\documentclass[aps,pre,reprint,
 twocolumn,
 notitlepage,
 showpacs,floatfix,nofootinbib,
 superscriptaddress
 ]{revtex4-1}

 \usepackage{subfigure}
 \usepackage{amssymb}
 \usepackage{amsfonts}
 \usepackage{amsmath}
 \usepackage{amsthm}
 \usepackage{epsfig}
 \usepackage{float}
 \usepackage[usenames,dvipsnames]{color}
 \usepackage[latin1]{inputenc}

 \usepackage{graphics,graphicx,color,subfigure}

 \usepackage[hidelinks,unicode=true]{hyperref}
 \hypersetup{colorlinks=true,
 	linkcolor=blue,
 	urlcolor=blue,
 	citecolor=blue,
 	pdfhighlight=/N
 }
 
 \usepackage{comment}
 \usepackage[normalem]{ulem}

\newcommand{\av}[1]{\langle {#1} \rangle}

\newcommand{\Ninf}{N_\mathrm{inf}}
\newcommand{\Ne}{N_\mathrm{e}}
\newcommand{\NSI}{N_\mathrm{SI}}

\newcommand{\tauinf}{\tau_{kk'}^\text{(inf)}}
\newcommand{\kmax}{k_\text{max}}
\newcommand{\kmin}{k_\text{min}}
\newcommand{\modthre}{\mathcal{T}} 
\newcommand{\modall}{\mathcal{A}}  
\newcommand{\modstan}{\mathcal{S}}  

\begin{document}

\title{Robustness and fragility of the susceptible-infected-susceptible epidemic models on complex networks}

\author{Wesley Cota}
\affiliation{Departamento de F\'{\i}sica, Universidade Federal de Vi\c{c}osa, 36570-900 Vi\c{c}osa, Minas Gerais, Brazil}

\author{Ang\'{e}lica S. Mata }
\affiliation{Departamento de F\'{\i}sica, Universidade Federal de Lavras, 37200-000, Lavras, Minas Gerais, Brazil}

\author{Silvio C. Ferreira}
\affiliation{Departamento de F\'{\i}sica, Universidade Federal de Vi\c{c}osa, 36570-900 Vi\c{c}osa, Minas Gerais, Brazil}
\affiliation{National Institute of Science and Technology for Complex Systems, Brazil}

\begin{abstract}
We analyze two alterations of the standard susceptible-infected-susceptible
(SIS) dynamics that preserve the central properties of spontaneous healing and
infection capacity of a vertex increasing unlimitedly with its degree.
{All models have the same epidemic thresholds in mean-field theories but} depending on the network
properties, simulations yield a dual scenario, in which the epidemic thresholds
of the modified SIS models can be either dramatically altered or remain
unchanged in comparison with the standard dynamics. For uncorrelated synthetic
networks having a power-law degree distribution with exponent $\gamma<5/2$, the
SIS dynamics are robust exhibiting essentially the same outcomes for all
investigated models. A threshold in better agreement with the heterogeneous
rather than
quenched mean-field theory is observed in the modified dynamics for exponent
$\gamma>5/2$. Differences are more remarkable for $\gamma>3$ where a finite
threshold is found in the modified models in contrast with the  vanishing
threshold of the original one. This duality is elucidated in terms of
{epidemic lifespan on star graphs}. We verify that the activation 
{of} the modified SIS models is triggered in the
innermost component of the network given by a $k$-core decomposition for
$\gamma<3$ while  it happens only for $\gamma<5/2$ in the standard model. For
$\gamma>3$, the activation in  the modified dynamics is collective involving
essentially the whole network  while it is triggered by hubs in the standard
SIS. The duality also appears in the finite-size scaling of the critical
quantities where mean-field behaviors are observed for the modified, but not for
the original dynamics. Our results feed the discussions about the most proper
conceptions of epidemic models to describe real systems and the choices of the
most suitable theoretical approaches to deal with these models.
\end{abstract}


\maketitle

\section{Introduction}
\label{sec:intro}

Network science has been marked by its interdisciplinary nature since its
consolidation as a new branch~\cite{barabasi2016network,Albert02}, especially
the investigation of  dynamical processes on networked
substrates~\cite{barratbook}. Epidemic spreading, one of the most prominent and
widely investigated issues, is usually investigated  by means of stochastic agent-based
 models~\cite{PSRMP}. Despite several advances in the understanding
of epidemic models on
networks~\cite{PSRMP,Chatterjee09,Durrett10,Castellano2010,Gleeson2013,Kitsak2010,newman02,Boguna2013}, it  remains target of recent intensive investigations~\cite{sander2016,Mata2015,DeArruda2017,St-Onge2017,Cai2016a,Wei2017,Castellano2017,Chen2018}.

One of the most basic but still not fully understood epidemic processes on
networks is the susceptible-infected-susceptible (SIS) model~\cite{PSRMP}, which
consists of agents lying on the vertices of a network which can be infected
or susceptible. Infected
individuals  become spontaneously healed (susceptible) with rate $\mu$ and
transmit the disease to their susceptible contacts with rate $\lambda$.  In
principle, the SIS dynamics can exhibit a phase transition between a disease-free
(absorbing) state and an active stationary phase, in which the epidemics
persists in an endemic state. The transition occurs at an epidemic threshold
$\lambda_c$. However, for  uncorrelated random networks with a power-law
degree distribution $P(k)\sim k^{-\gamma}$, it was rigorously
proved~\cite{Chatterjee09} and later put in sound physical
grounds~\cite{Boguna2013} that the absorbing phase is unstable in the
thermodynamic limit implying that the epidemic threshold is formally zero.

Considering that both real and computationally generated networks are finite,
the finite-size dependence of the epidemic variables is a fundamental issue.
Analytically, it is frequently accessed by mean-field approximations that take
into account the network heterogeneity,  but truncate at some level the
dynamical correlations~\cite{Gleeson2013}. Two classes of mean-field theories
are mostly used. The degree-based theory~\cite{Pastor01,Pastor01b}, termed as
heterogeneous mean-field (HMF)~\cite{barabasi2016network,barratbook}, is a
coarse-grained mixing approach, in which the vertex degree is the relevant
quantity. This method is closely related to the annealed network regime where
the connections are rewired in time scales much shorter than those of the
dynamical processes taking place on the top of the
network~\cite{Dorogovtsev08,PSRMP}. The individual-based
theory~\cite{Wang03,VanMieghem2012,Chakrabarti08}, termed quenched mean-field
(QMF)~\cite{Castellano2010}, considers the network structure without mixing
using its adjacency matrix~\cite{barabasi2016network}. These theories predict
equivalent epidemic thresholds of the SIS dynamics on uncorrelated random
networks with power-law degree distribution of exponent $2<\gamma<5/2$ but are
sharply conflicting for $\gamma>3$~\cite{Castellano2010}, for which HMF predicts
finite thresholds whereas QMF vanishing ones as $N\rightarrow\infty$. The latter
is asymptotically  in agreement with the exact
results~\cite{Durrett10,Chatterjee09} and supported by stochastic
simulations~\cite{Ferreira12,mata2013pair,Boguna2013}. For $5/2<\gamma<3$, both
theories state a null threshold as $N\rightarrow\infty$ but the way that the
asymptotic value is approached and, thus, the effective finite-size thresholds
are different. Improvements of these theories including dynamical correlations
by means of pairwise approximations~\cite{Gleeson2013} do not change the
foregoing scenarios~\cite{mata2013pair,Mata14,Cai2016a}.

Recently, a criterion formerly conceived for SIS model~\cite{Boguna2013} was
applied to determine the nature of epidemic thresholds of generic processes on
networks with power-law degree distributions~\cite{sander2016}. The criterion
involves the recovering time $\tau_k$ of an epidemics on a star graph,
consisting of a central vertex connected to $k$ leaves of degree 1 that mimics
the hubs of a network, and the time $\tau^\text{(inf)}$ that the hubs take to
mutually transmit the infection to each other. If $\tau_k \gg
\tau^\text{(inf)}$, hubs remain active for times sufficiently long to infect
each other and the epidemics is triggered by the mutual activation of hubs,
leading to a vanishing threshold in the thermodynamic limit. If $\tau_k \lesssim
\tau^\text{(inf)}$, the mutual reinfection is knocked out and the transition to
an endemic phase can only take place collectively involving a finite fraction of
the network and happens at a finite threshold. In Ref.~\cite{sander2016}, this
criterion notably predicted that waning immunity~\cite{anderson92}, in which
infected individuals are temporarily immunized before to become susceptible,
leads to a finite threshold for $\gamma>3${ which disagrees with the
	prediction of QMF approximation},
but in agreement with extensive numerical simulations.

A fundamental question naturally arises. How robust is the hub mutual activation
mechanism of the standard SIS dynamics? In the present work,  we tackle this
problem comparing slightly modified versions of the standard SIS model,
preserving the  spontaneous healing and infection capacity increasing
proportionally to the vertex degree. All modified and original models have the
same thresholds in both HMF and QMF theories. However, the criterion of mutual
reinfection time of hubs~\cite{sander2016,Boguna2013} predicts a finite
threshold in the thermodynamic limit for the modified models in 
uncorrelated networks~\cite{Catanzaro05} with $\gamma>3$, in contrast with the
standard SIS. Stochastic simulations~\cite{Cota} on large  networks
corroborate this prediction. For $5/2<\gamma<3$, we observed that the modified
dynamics present a vanishing threshold in better agreement with HMF than QMF.
For $2<\gamma<5/2$, the SIS infection mechanism is robust and  all models have
essentially the same epidemic threshold. This duality is explained in terms of
epidemic activation mechanisms~\cite{Kitsak2010,Castellano12,sander2016}.

Our results gathered with  previous reports of Ref.~\cite{sander2016}, in which
waning immunity can drastically change the threshold behavior, lead to the
following take-home messages. Firstly, the metastable, localized, and active
states of the standard SIS dynamics necessary to sustain the endemic activity
for any infection rate for $\gamma>3$ are not universal and their realizations
in real epidemic processes may be unrealistic. Second, for the widely more
frequent case of networks with $2<\gamma<3$, the null threshold is a robust
feature, obtained irrespectively of the existence of locally self-activated star
subgraphs. In such an absence, epidemics is triggered in the innermost, densely
connected component of the network given  by a $k$-core
decomposition~\cite{Dorogovtsev2006a}, while for the original SIS model it
happens only for $2<\gamma<5/2$~\cite{Castellano12}. Last but not least, the
HMF theory~\cite{Pastor01,barratbook,PSRMP}, which has
been frequently pretermitted due to its failure in capturing the asymptotically
null epidemic threshold of the standard SIS for
$\gamma>3$~\cite{Wang03,Castellano2010,VanMieghem2012},
{ is more accurate than} the QMF theory for the {present} modified SIS models 
and also in other models as contact
processes~\cite{Mata14} and SIRS~\cite{sander2016}. The origins of this
worse performance of the QMF theory is discussed in our
conclusions.

The remaining of the paper is organized as follows. Section~\ref{sec:models}
describes the investigated models, and their mean-field theories are discussed
in Sec.~\ref{sec:mf}. Epidemic thresholds obtained in numerical simulations
are presented and compared with mean-field theories in Sec.~\ref{sec:ucm}.
The finite-size scaling of the critical quantities are provided in
Sec.~\ref{sec:fss}. We draw our concluding remarks {and prospects} in
Sec.~\ref{sec:conclu}. Appendices~\ref{app:simu},  \ref{app:pair}, and
\ref{app:lsanalytic}  complement the
paper with analytical and numerical  details.

\section{Epidemic models}
\label{sec:models}

We investigate three epidemic dynamics where each vertex of the network
can be either infected or susceptible. The infected ones are spontaneously
healed with rate $\mu$ in all models. In the \textit{standard} SIS, hereafter
called SIS-$\modstan$, an infected vertex infects each susceptible
nearest-neighbor with rate $\lambda$. In the SIS-$\modthre$ model, infection
is a \textit{threshold} process where susceptible vertices are infected with rate
$\lambda$ if they have at least one infected nearest-neighbor\footnote{This is
	an asynchronous version of the model investigated in seminal
	papers~\cite{Pastor01a,Pastor01} dealing with epidemic spreading on networks.}.
Finally, SIS-$\modall$ is a modification of the contact process~\cite{Marrobook}
where the infected vertices simultaneously infect \textit{all} susceptible
neighbors with  rate $\lambda$. The symbols $\modstan$, $\modthre$, and
$\modall$ make reference to \textit{standard}, \textit{threshold}, and
\textit{all} in the model definitions.   The  models rules and some details of
their computer implementations described in Appendix~\ref{app:simu} are
summarized in Table~\ref{tab:rules}.

\begin{table*}
\centering
\def\arraystretch{1.5}
\caption{Epidemic model definitions and some computer implementation details of
	the Gillespie algorithm (GA) presented in Appendix~\ref{app:simu}. Symbols:
		$\Ninf$ is the number of infected vertices; $\NSI$ is number of susceptible
		vertices with at least one infected nearest-neighbor; $\Ne$ is the number of edges
		emanating from infected vertices; and $u$ is random variable uniformly
		distributed in the interval $(0,1)$. }
\label{tab:rules}
\begin{tabular}{ccccc}
\toprule
SIS-$\modthre$ (\textit{threshold})~\cite{Dickman1988,Bottcher2018} & ~~ & SIS-$\modall$ (\textit{all})~\cite{Sander2009,DeOliveira2016} & ~~ &    SIS-$\modstan$ 
(\textit{standard})~\cite{PSRMP} \\\hline
Infected vertices are       & ~~ & Infected vertices are    & ~~ & Infected vertices are  \\
spontaneously  healed       & ~~ & spontaneously healed     & ~~ & spontaneously healed   \\
with rate $\mu$             & ~~ & with rate $\mu$          & ~~ &   with rate $\mu$      \\ 
                            & ~~ &                          & ~~ &                        \\ 
Susceptible vertices become & ~~ & Infected vertices infect &  ~~ & Infected vertices      \\
infected with rate $\lambda$& ~~ & at once all susceptible  & ~~ & independently infect \\
if they have at least one   & ~~ &   neighbors with         & ~~ & each susceptible neighbor \\
infected neighbor   & ~~ &    rate $\lambda$        & ~~ & with rate $\lambda$ \\
                            & ~~ &                          & ~~ &                        \\ 
GA infection probability & ~~ & GA infection probability    & ~~ & GA infection probability\\ 
$q=\dfrac{\lambda \NSI}{\mu\Ninf+\lambda \NSI}$& ~~ & $q=\dfrac{\lambda}{\mu+\lambda}$ & ~~ &  $q=\dfrac{\lambda\Ne}{\mu\Ninf+\lambda\Ne}$                       \\ 
                            & ~~ &                          & ~~ &                        \\ 
GA time step                & ~~ & GA time step          & ~~ & GA time step \\ 
$\tau = \dfrac{-\ln(u)}{\mu\Ninf+\lambda \NSI}$& ~~ & $\tau=\dfrac{-\ln(u)}{(\mu+\lambda)\Ninf}$ & ~~ &  $\tau=\dfrac{-\ln(u)}{\mu\Ninf+\lambda\Ne}$                       \\ 
\toprule
	\end{tabular}
\end{table*}
The modified dynamics preserve two central features of the standard SIS model:
spontaneous healing and infection capacity of a vertex increasing
proportionally to its degree. All  models have their counterparts in
regular lattices with a fixed coordination number $k$: SIS-$\modstan$ can be
mapped in the contact process (CP)~\cite{Harris,Marrobook}, in which infected
vertices transmit to a nearest-neighbor chosen at random with rate
$\lambda_\text{CP}$ and heals spontaneously, using
$\lambda_\text{SIS}=\lambda_\text{CP}/k$. SIS-$\modall$ was investigated in
Refs.~\cite{Sander2009,DeOliveira2016} while SIS-$\modthre$ was investigated in
Refs.~\cite{Dickman1988,Bottcher2018}. In lattices, all models belong to the
directed percolation universality class~\cite{Marrobook}.

\begin{figure}[bth]
	\centering
	\includegraphics[width=0.65\linewidth]{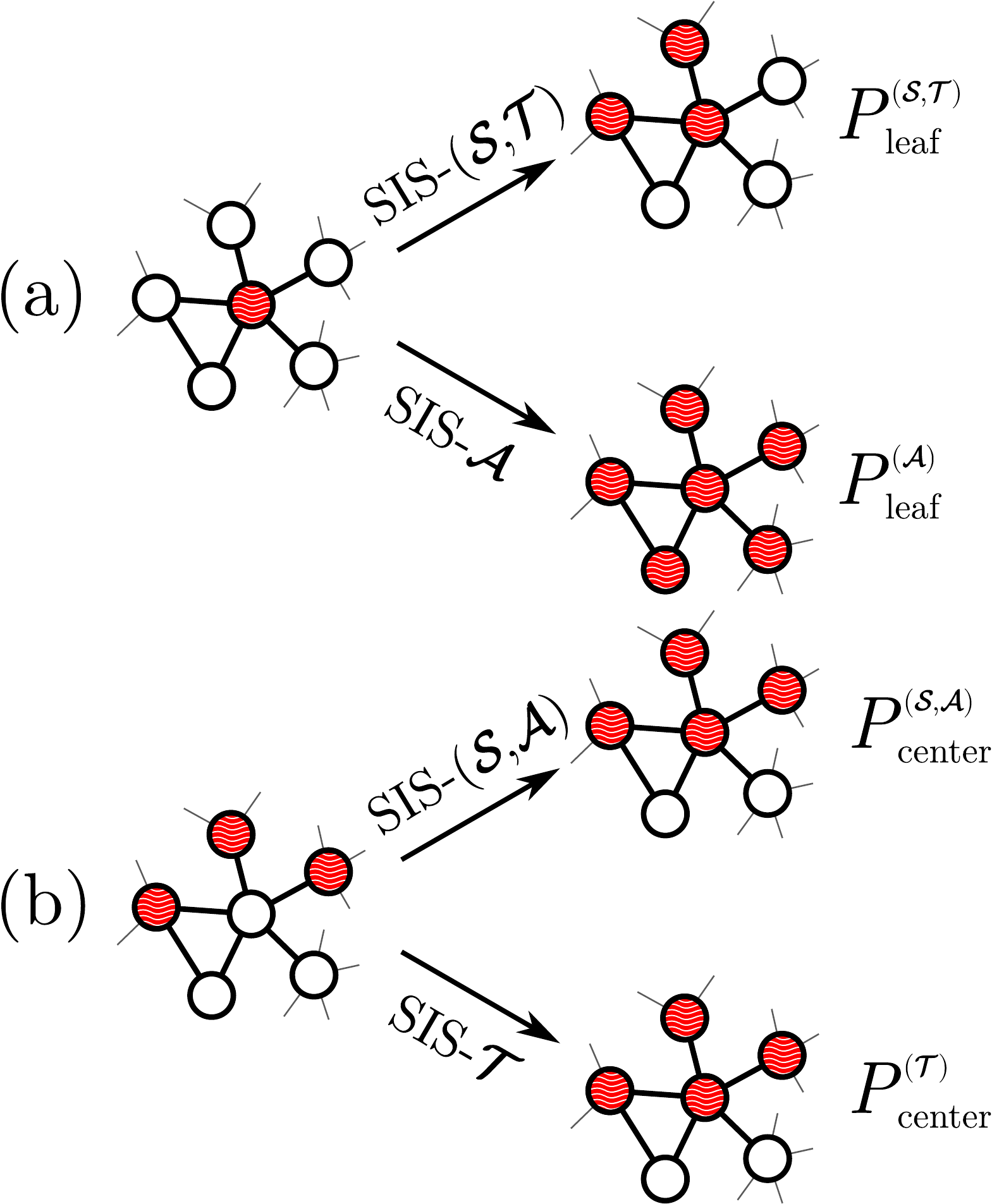}
	\caption{Some infection processes in the SIS models. (a) An infected
		vertex (center) with many susceptible neighbors (leaves). (b) A
		susceptible center with  infected leaves. Transition probabilities
		are defined in Eqs.~\eqref{eq:Psaleaf}--\eqref{eq:Pacenter}.}
	\label{fig:models}
\end{figure}

Figure~\ref{fig:models} shows two important situations where the modified models
differ from the standard SIS. Consider an infinitesimal time interval $\Delta t$
and an infected vertex (the center) surrounded by $k$ susceptible neighbors
(leaves); see Fig.~\ref{fig:models}(a). The probability that $s$ leaves are
infected by the center for both SIS-$\modstan$ and SIS-$\modthre$ is
\begin{equation}
P^{(\modstan,\modthre)}_\text{leaf}(s)=\binom{k}{s}(\lambda \Delta t)^s(1-\lambda \Delta t)^{k-s},
\label{eq:Psaleaf}
\end{equation}
while for SIS-$\modall$ it is
\begin{equation}
P^{(\modall)}_\text{leaf}(s)=\lambda \Delta t \delta_{s,k},
\label{eq:Pbleaf}
\end{equation}
where $\delta_{s,k}$ is the Kronecker delta symbol. Note that both expressions
produce the same mean number of infected leaves $\av{s}=\lambda k
\Delta t$. Now, the probability that a susceptible center
surrounded by $s>0$ infected leaves,  Fig.~\ref{fig:models}(b), is
infected is given by
\begin{equation}
P^{(\modstan,\modall)}_\text{center}(s) =  1-(1-\lambda\Delta t)^s\approx \lambda s \Delta t
\label{eq:Psbcenter}
\end{equation} 
for  SIS-$\modstan$ and SIS-$\modall$ while for SIS-$\modthre$
it becomes 
\begin{equation}
P^{(\modthre)}_\text{center}=\lambda\Delta t.
\label{eq:Pacenter}
\end{equation} 
So, while the infection of leaves by the center  in SIS-$\modstan$ is
equivalent to SIS-$\modthre$, the infection of the center by leaves in SIS-$\modstan$ is
equivalent to SIS-$\modall$.

The simulations of these models were performed using the algorithms described in
Appendix~\ref{app:simu}, which include phantom processes~\cite{Cota} in the
statistically exact Gillespie algorithm (GA)~\cite{Gillespie1976403} for the
simulations of general Markovian stochastic processes. Some important
implementation details are highlighted in Table~\ref{tab:rules}. The equivalence
between  optimized prescriptions and the original GA as well as their
computational performances for several models, including SIS-$\modstan$, can be
found in Ref.~\cite{Cota}. The implementations of SIS-$\modthre$ and $\modall$ 
can be derived in an analogous way.

\section{Mean field analysis} 
\label{sec:mf}

The HMF theory consists in dynamical equations for the probability $\rho_k$ that
a vertex of degree  $k$ is infected and disregards the stochasticity of the process. 
The probability that a neighbor of a vertex of degree $k$ is infected reads
as~\cite{Pastor01a} $\Theta_k =\sum_{k'}P(k'|k)\rho_{k'}$ where $P(k'|k)$ is the
probability that a neighbor of a vertex  with degree $k$ has degree $k'$. So,
$\rho_k$ evolves as
\begin{equation}
\frac{d \rho_k}{d t}=-\mu \rho_k+\lambda (1-\rho_k)\Psi_k(\Theta_k)
\label{eq:HMFgen}
\end{equation}
where $\Psi_k(\Theta_k)=k\Theta_k$ for SIS-$\modstan$ and SIS-$\modall$, and
$\Psi_k(\Theta_k)=1-(1-\Theta_k)^k$ for SIS-$\modthre$. The QMF theory consists of
dynamical equations for the probability $\rho_i$ that a vertex  $i$ is infected
and reads as
\begin{equation}
\frac{d\rho_i}{d t} = -\mu \rho_i+\lambda (1-\rho_i) \Psi_i,
\label{eq:QMFgen}
\end{equation}
where $\Psi_i=\sum_{j}A_{ij}\rho_j$ for SIS-$\modstan$ and SIS-$\modall$, and
\[\Psi_i=1-\prod_{j|A_{ij}=1} (1-\rho_j)\] 
for SIS-$\modthre$, in which and the adjacency matrix is given by $A_{ij}=1$ if $i$ and $j$
are connected and $A_{ij}=0$ otherwise. The multiple simultaneous infections in
SIS-$\modall$ do not play a role in these one-vertex mean-field theories since there
are no multiple connections. It is worth to mentioning that the HMF theory of
SIS-$\modthre$ for uncorrelated networks {with} $P(k'|k)=
k'P(k')/\av{k}$~\cite{vazquez}  was recently investigated~\cite{Morita2015}.

The mean-field epidemic thresholds can be obtained   with  the stability
analysis and linearization of Eqs.~\eqref{eq:HMFgen} and \eqref{eq:QMFgen} 
around the fixed points $\rho_k=0$ or $\rho_i=0$, respectively. The linearized equations
are the same for the three models 
\begin{equation}
	\frac{d\rho_k}{dt} = -\mu \rho_k+\lambda\sum_{k'}C_{kk'}\rho_{k'}
\end{equation}
and
\begin{equation}
\frac{d\rho_i}{dt} = -\mu \rho_i+\lambda\sum_{j}A_{ij}\rho_j,
\end{equation}
where $C_{k'k}=kP(k'|k)$. The HMF and QMF thresholds are obtained
when the largest eigenvalue of the respective Jacobians 
$J_{kk'}^\text{HMF}=-\mu\delta_{kk'}+\lambda C_{kk'}$ and  
$J_{ij}^\text{QMF}=-\mu\delta_{ij}+\lambda A_{ij}$ are zero.
For the HMF theory, it is given by~\cite{PREboguna}
\begin{equation}
\lambda_c^\text{HMF}=\frac{1}{\Upsilon_\text{max}}
\label{eq:thr_HMFgen}
\end{equation}
 where $\Upsilon_\text{max}$ is the largest eigenvalue of $C_{k'k}$.
 For uncorrelated networks we obtain 
 \begin{equation}
\lambda_c^\text{HMF}=\frac{\av{k}}{\av{k^2}}
\label{eq:thr_HMFann}
 \end{equation}
 where $\av{k^s}=\sum_k k^sP(k)$.
 For  the QMF theory, we obtain~\cite{Castellano2010}
 \begin{equation}
 \lambda_c^\text{QMF}=\frac{1}{\Lambda_\text{max}}
 \label{eq:thr_QMFgen}
 \end{equation}
 where $\Lambda_\text{max}$ is the largest eigenvalue of the adjacency matrix $A_{ij}$.

\begin{figure}[hbt]
	\centering
	\includegraphics[width=0.95\linewidth]{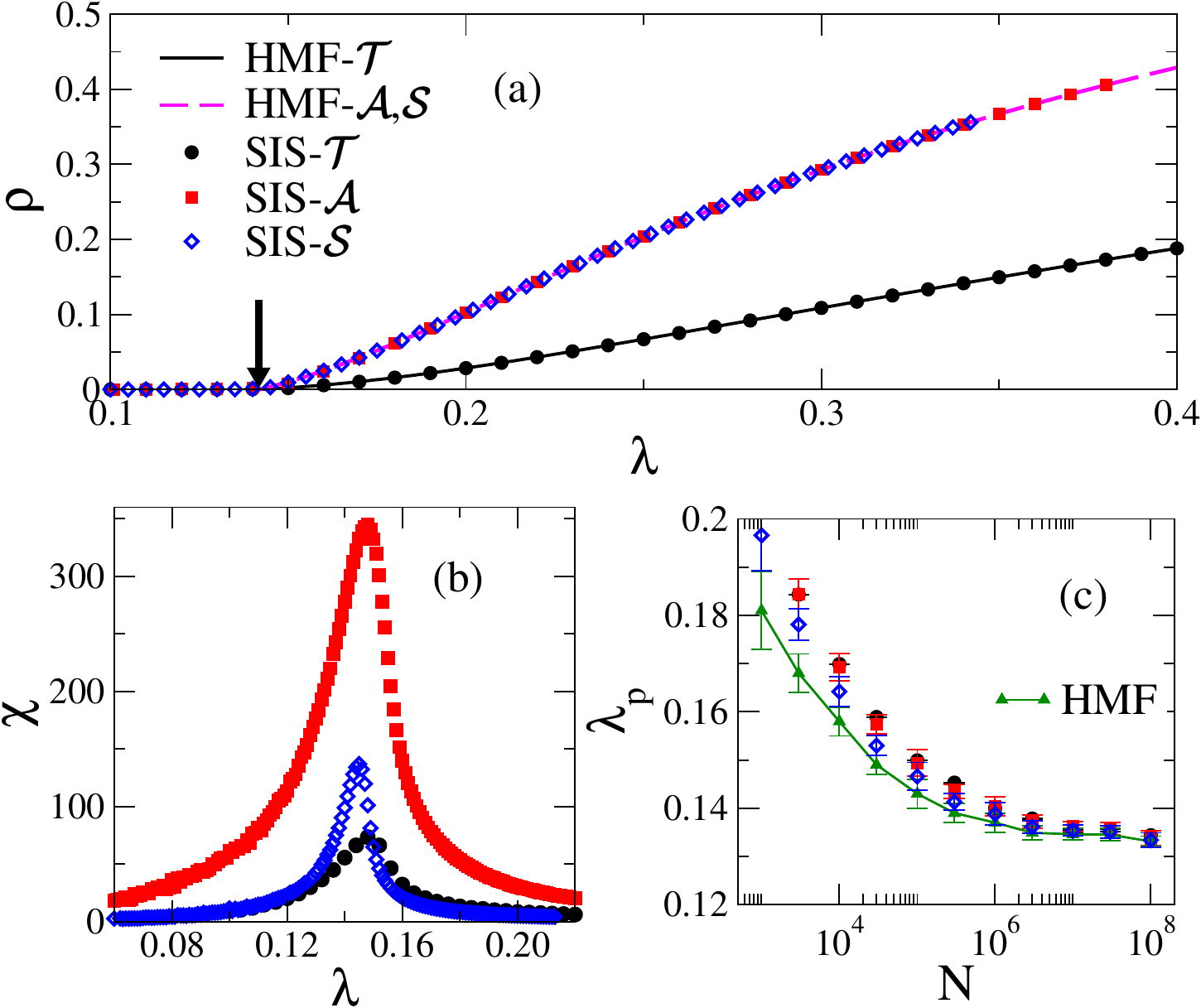}
	\caption{Comparison of HMF theory and simulations on annealed networks with
		$N=10^5$ vertices, degree distribution $P(k)\sim k^{-3.5}$, minimal degree
		$k_\text{min}=3$, and upper cutoff $k_\text{c}=\sqrt{N}$. (a) QS density and (b)
		susceptibility versus infection rate curves are shown. Lines in (a) are numerical
		solutions of Eq.~\eqref{eq:HMFgen} in the stationary regime and the arrow
		indicates the HMF epidemic threshold $\lambda_c^\text{HMF}=\av{k}/\av{k^2}$. 
		(c) Finite-size dependence of the threshold estimated via susceptibility and
		HMF theory. The curves correspond to averages over 10 independent network
		realizations.}
	\label{fig:hmfanng3p5}
\end{figure}
\begin{figure*}[hbt!]
	\centering
	\includegraphics[width=0.75\linewidth]{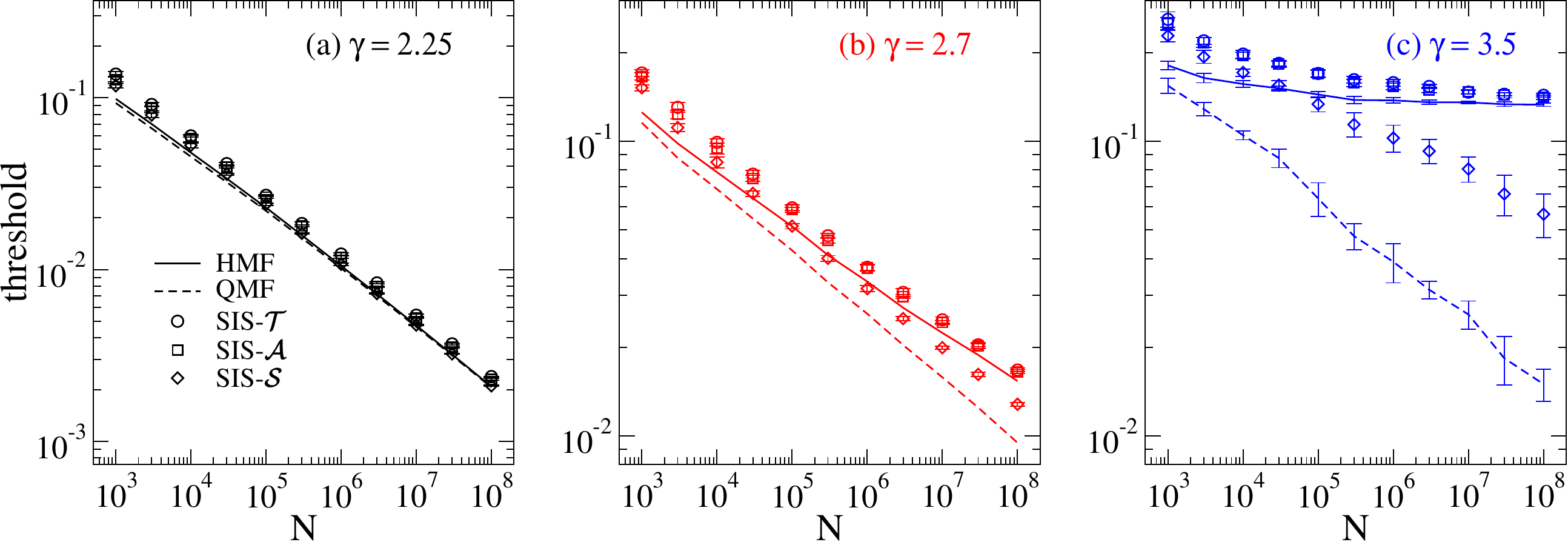}
	\caption{Epidemic thresholds for SIS models on UCM networks with
		$k_\text{min}=3$, $k_\text{c}=\sqrt{N}$, and different degree exponents (a)
		$\gamma=2.25$, (b) $2.7$, and (c) $3.5$. Solid and dashed lines correspond to
		HMF and QMF theories, respectively.  Curves are averages over 10 network
		realizations. Negligible error bars in mean-field theories are not shown.}
	\label{fig:threshold_ucm}
\end{figure*}
The HMF theory on uncorrelated networks was compared  with  the numerical
simulations on annealed networks (see Appendix~\ref{app:simu}), for which this
theory exactly predicts the threshold and average density of infected vertices
in the thermodynamic limit~\cite{Dorogovtsev08,Boguna09}. Simulations with
absorbing states near the transition need special techniques~\cite{Cota}. We use
here the standard quasistationary (QS) method described in Ref.~\cite{SanderQS},
in which the averaging is constrained to the active states  and converges to the
actual stationary phase in the thermodynamic limit. The threshold in finite
networks can be estimated using the principal peak of the dynamical
susceptibility $\chi$ defined in the QS state as~\cite{Ferreira12}
\begin{equation}
\chi =N\frac{\av{\rho^2}-\av{\rho}^2}{\av{\rho}}.
\label{eq:sus}
\end{equation}

Figures~\ref{fig:hmfanng3p5}(a) and (c) confirm the agreement between
simulations on annealed networks and HMF theory for the stationary densities and
the thresholds, respectively, in all investigated models. However, the
fluctuations of the order parameter
are different as shown by the susceptibility curves in Fig.~\ref{fig:hmfanng3p5}(b).
See also Sec.~\ref{sec:fss}.

\section{SIS models on  synthetic quenched networks}
\label{sec:ucm}

\subsection{Epidemic thresholds}

We investigate networks having power-law degree distribution $P(k) \sim
k^{-\gamma}$, generated with the uncorrelated configuration model
(UCM)~\cite{Catanzaro05} with minimal vertex degree $k_\text{min}=3$ and
structural upper cutoff $k_\text{c}=\sqrt{N}$, granting the absence
of degree correlations~\cite{mariancutofss} permitting, therefore, comparison
with the HMF epidemic threshold given by Eq.~\eqref{eq:thr_HMFann}. The
thresholds obtained in simulations are compared with HMF and QMF theories in
Fig.~\ref{fig:threshold_ucm}. 

For $\gamma<5/2$, here represented by $\gamma=2.25$ in
Fig.~\ref{fig:threshold_ucm}(a), all models have approximately the same
threshold well described by both HMF and QMF theories, which have already been
reported for SIS-$\modstan$~\cite{Ferreira12}.

For $5/2<\gamma<3$, represented by $\gamma=2.7$ in
Fig.~\ref{fig:threshold_ucm}(b), SIS-$\modthre$ and SIS-$\modall$ have
essentially the same threshold whose scaling is very well fitted by the HMF theory
and deviates from QMF. The threshold of the standard SIS-$\modstan$ vanishes
with a scaling deviating from both HMF and QMF scalings. A good agreement
between  the  threshold of the standard SIS for $\gamma=2.7$ can be recovered
with the pairwise QMF theory of Ref.~\cite{mata2013pair} but not with the
pairwise HMF theory of Refs.~\cite{Mata14,Cai2016a}; see Appendix~\ref{app:pair}.

The results for modified SIS models are markedly contrasting with the standard
one\footnote{In the case of multiple peaks, which can be observed in SIS-$\modstan$
	on large UCM networks with $\gamma>3$~\cite{Ferreira12,mata2013pair}, the
	principal peak is the one that provide a threshold closest to the lifespan
	divergence and matches the threshold of the lifespan  method proposed in
	Ref.~\cite{Boguna2013}; see Ref.~\cite{Mata2015}.} for $\gamma>3$, represented
by $\gamma=3.5$ in Fig.~\ref{fig:threshold_ucm}(c). The modified
SIS-$\modthre$ and $\modall$ dynamics present a finite threshold in very satisfactory
accordance with  HMF theory and contrasting with the original SIS-$\modstan$
that presents the well-known threshold approaching zero as the size increases. Note,
however, that the thresholds of SIS-$\modstan$  have a scaling incompatible with
QMF for the investigated size range that cannot be reckoned by neither
pairwise QMF~\cite{mata2013pair} or HMF~\cite{Mata14,Cai2016a} theories (see
Appendix~\ref{app:pair}). The latter still predicts a finite threshold,
inconsistent with simulations and the rigorous results~\cite{Chatterjee09} for
SIS-$\modstan$.
\begin{figure}[hbt]
	\centering
	\includegraphics[width=0.68\linewidth]{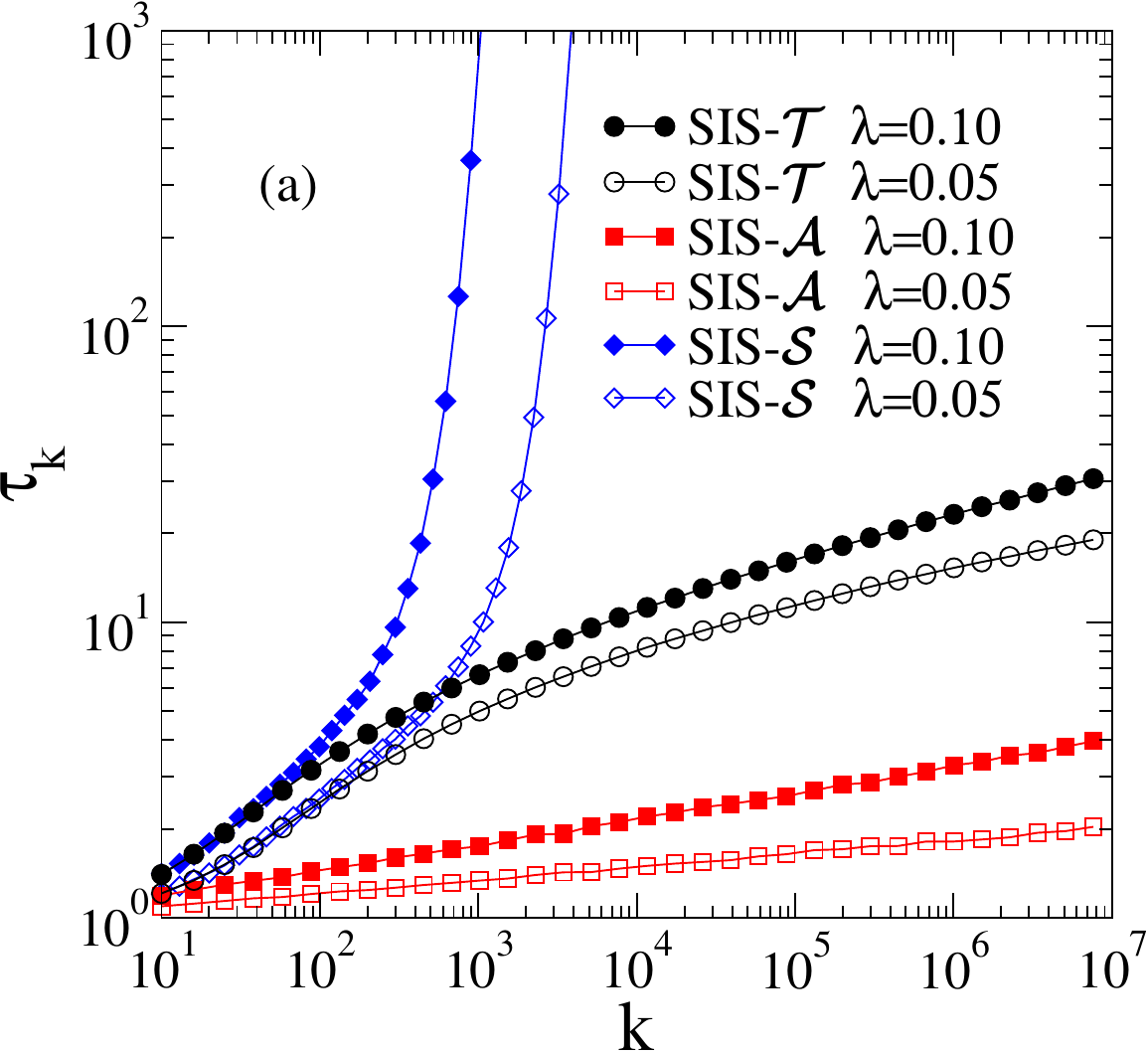}\\
	\includegraphics[width=0.68\linewidth]{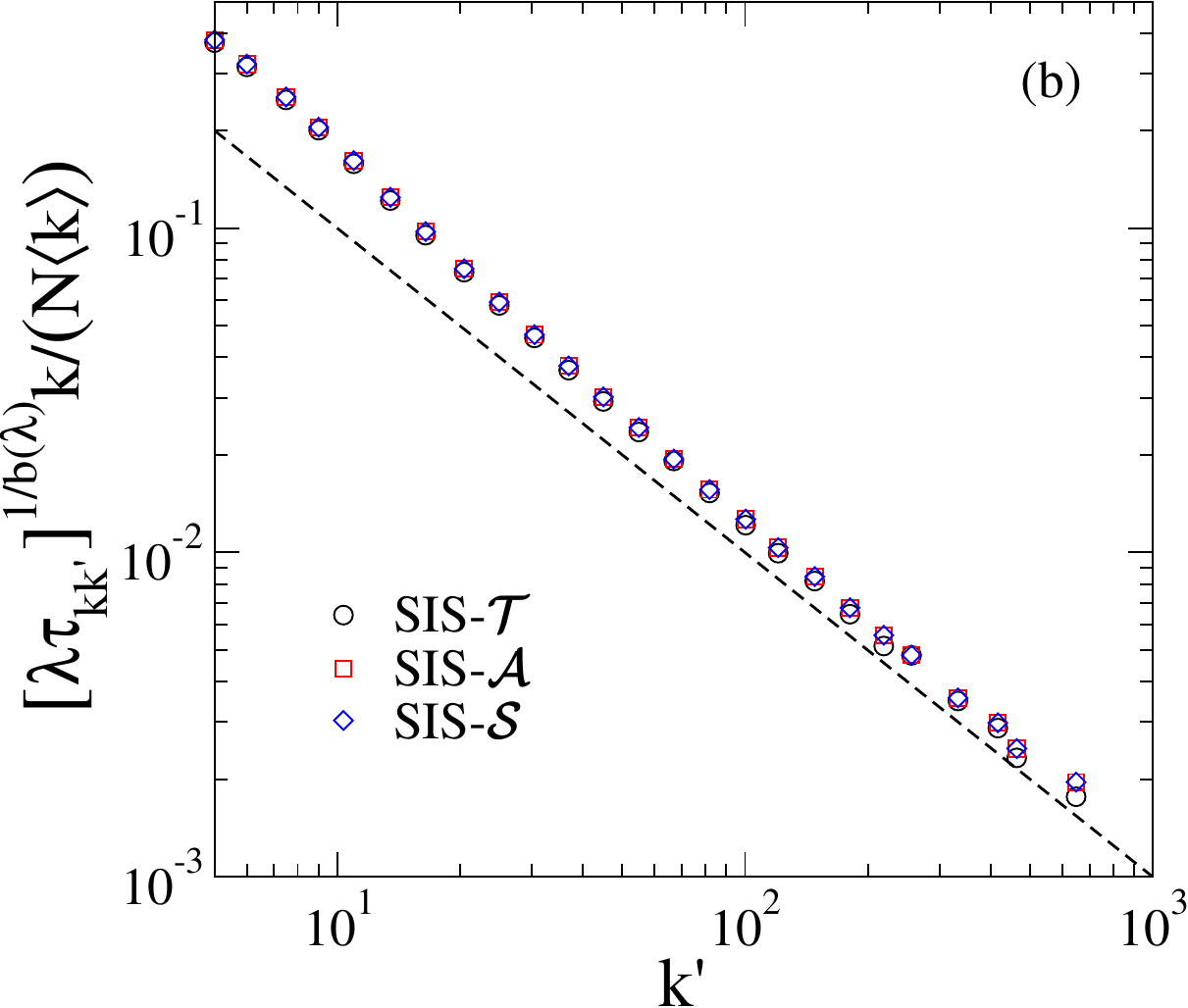}
	\caption{(a) Activity lifespan for epidemic processes on star graphs. The
		initial condition is the center infected and all leaves susceptible. The number
		of runs varies from $10^3$ to $10^5$, the larger {number} the smaller $\lambda$.
		(b) Mutual reinfection of hubs scaled according to Eq.~\eqref{eq:taukk}. The
		degree exponent is $\gamma=3.5$, the size is $N=10^6$ and infection rate is
		$\lambda=0.05$. The vertex that is kept infected has degree $k=50$. The dashed line is the prediction of the right-hand side of Eq.~\eqref{eq:taukk}. }
	\label{fig:starlifespan}
\end{figure}

\subsection{Activation mechanisms for $\gamma>3$}

To clarify the antagonistic results for $\gamma>3$, we consider the
recovering time of the epidemics  on star graphs for small values of $\lambda$.
Figure~\ref{fig:starlifespan}(a) shows the epidemic lifespan for the distinct
SIS models as a function of the star graph size. For standard SIS-$\modstan$, we
see an exponential growth predicted by the approximated discrete time dynamics 
of Ref.~\cite{Boguna2013} (also Ref.~\cite{sander2016}) given by $\tau_k^{(\modstan)}
\approx \frac{2}{\mu} \exp[k (\lambda/\mu)^2]$; see
Appendix~\ref{app:lsanalytic}. However, SIS-$\modthre$ and
SIS-$\modall$ present epidemic lifespans increasing very slowly with graph size,
consistent with a logarithmic growth. Applying the discrete time approach, a 
finite lifespan is obtained for SIS-$\modall$ and, after some refinement of the
theory, a logarithmic increase is found for SIS-$\modthre$; see
Appendix~\ref{app:lsanalytic} for details. {Indeed, the activities in
	SIS-$\modall$ are more correlated, and this has a significant effect on the
	probability of hub activation.}

An upper bound for the long-range infection times of hubs of degrees $k$ and
$k'$, denoted by $\tau^\text{(inf)}_{kk'}$, for uncorrelated networks can be
obtained following the same steps of Ref.~\cite{Boguna2013}
(also Ref.~\cite{sander2016}). The result is the same for all investigated SIS models
and given by
\begin{equation}
\tauinf\le \tau_{kk'}=\frac{1}{\lambda}
\left[\frac{N\av{k}}{kk'}\right]^{b(\lambda)}
\label{eq:taukk}
\end{equation}
where $b(\lambda)={\ln(1+\mu/\lambda)}/\ln\kappa$ and $\kappa =\av{k^2}/\av{k}$. Even being rigorously an upper bound, the right-hand side of
Eq~\eqref{eq:taukk} works very accurately  for $\lambda\ll \mu$  and $\gamma>3$
such that we can adopt $\tauinf\approx \tau_{kk'}$ as done for
SIS-$\modstan$~\cite{Boguna2013} and other epidemic models~\cite{sander2016}. This
agreement is confirmed in Fig.~\ref{fig:starlifespan}(b)
{for} $\gamma=3.5$. The simulation is run
keeping one single vertex of degree $k$  always infected (never heals) and
computing the time for the infection to reach for the first time each vertex of the
network, limited to a maximal time $t_\text{max}=10^{10}$. {Vertices that were not reached
	are not included in the averages but they represent a tiny fraction.}

With the approximation given by the right-hand side of Eq.~\eqref{eq:taukk}, we
have that $\tauinf \gtrsim \tau^\text{(inf)}_{\kmax,\kmax}$ where $\kmax$  is
the largest degree of the network that scales as $\av{\kmax}\sim
N^{1/(\gamma-1)}$ for UCM networks with $\gamma>3$~\cite{mariancutofss}. Also,
we have that $b(\lambda)$  is finite  since $\kappa$ converges to a constant as
$N\rightarrow\infty$ for $\gamma>3$, providing an algebraic increase of
$\tauinf$ with $N$. The condition
$\tau^\text{(inf)}\gg\tau_k^{(\modthre,\modall)}$ is obeyed such that epidemics
in the modified SIS models cannot be activated by hubs when $\lambda\ll\mu$ and
a collective phase transition at finite threshold is expected~\cite{sander2016}
in contrast with $\tau^\text{(inf)}\ll \tau_k^{(\modstan)}$ of the standard SIS,
in which the hub activation mechanism is at work and the threshold is null {in
	the thermodynamic limit}.

\subsection{Activation mechanisms for $2<\gamma<3$}

For $\gamma<3$, the hubs are sufficiently close~\cite{Hoyst2005}  to infect each
other even if their activity lifespans are not too large (exponential) and the
threshold goes to zero for all models as $N\rightarrow\infty$. However, there
exists a difference in the threshold scaling for $\gamma=2.7$ but does not for
$\gamma=2.25$. It has been claimed~\cite{Kitsak2010} that the most effective
spreaders in an epidemic processes lie  in a subset containing the innermost
core of the networks identified by the maximal index of the $k$-core
decomposition\footnote{A $k$-core decomposition consists of the following
	pruning process. Remove all vertices with degree $k_s=\kmin$ plus their edges
	and all other vertices that possess a degree $\kmin$ after the removal until no
	more vertices of degree $\kmin$ appear in the process. Next, the procedure is
	repeated for all vertices of degree $k_s=\kmin+1,~\kmin+2$ and so on until all
	vertices are removed. The maximal $k$-core  corresponds to the subset of
	vertices and edges removed in the last step of the decomposition.
}~\cite{Dorogovtsev08,Dorogovtsev2006a}. For SIS-$\modstan$, this mechanism is
claimed to hold for uncorrelated networks with $\gamma<5/2$ but the case
$5/2<\gamma<3$ has activation ruled by the hubs~\cite{Castellano12}. Since hubs
cannot be activated in isolation for arbitrarily small $\lambda$ in
SIS-$\modall$ and SIS-$\modthre$, we propose that the epidemic threshold should
be ruled by the subgraph identified by the maximal $k$-core for the whole range of
scale-free networks with $2<\gamma<3$.
\begin{figure}[hbt!]
	\centering
	\includegraphics[width=0.9\linewidth]{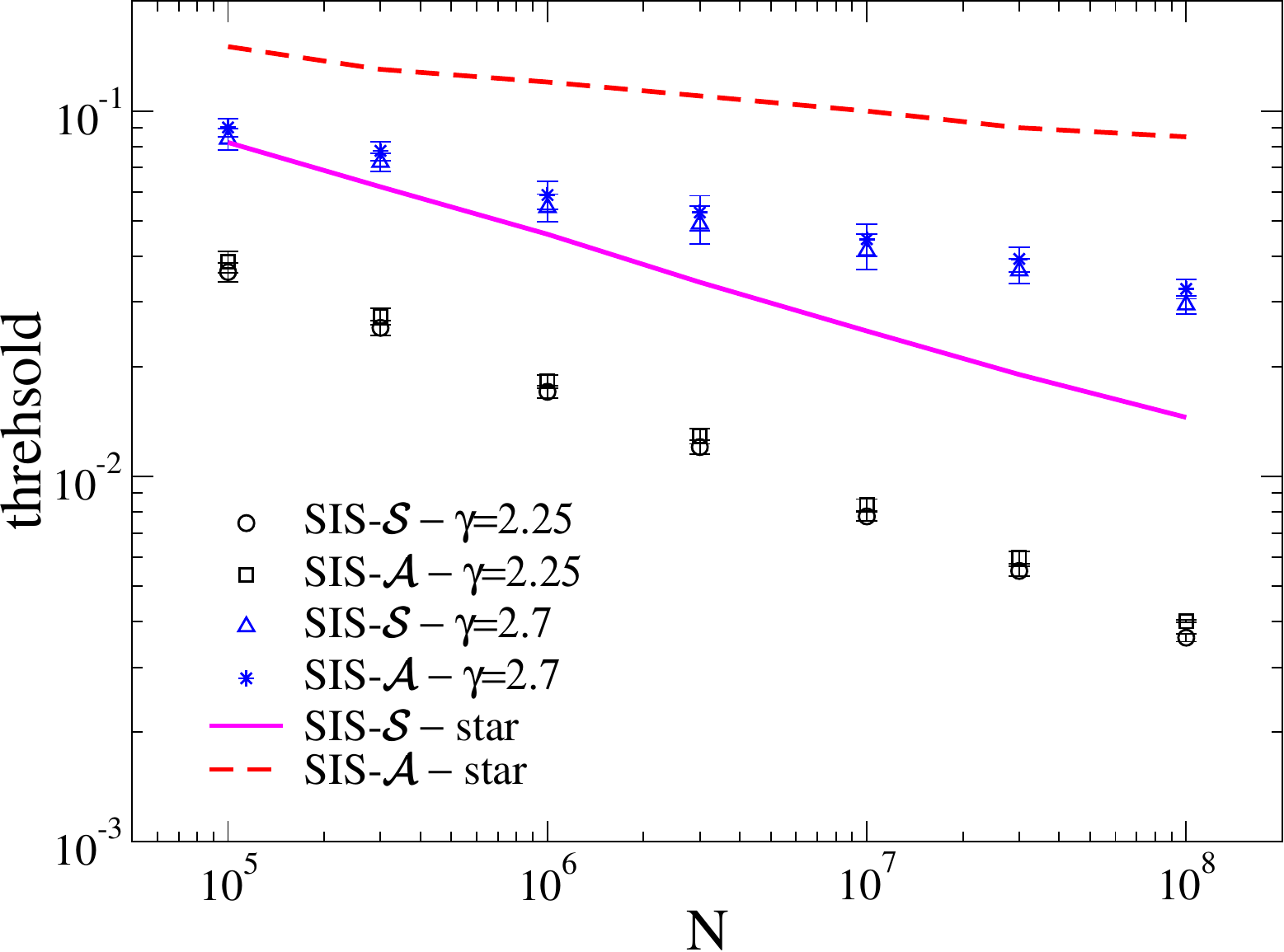}
	\caption{Epidemic thresholds for SIS-$\modstan$ and SIS-$\modall$
		running on the maximum $k$-core subgraph of networks with degree exponents
		$\gamma=2.25$ and $2.7$. The simulation results on star graphs with
		$\kmax\approx \sqrt{N}$ leaves are also presented. The averages were done over
		10 network realizations.
	}
	\label{fig:compare_sig_beta_kcore225and270}
\end{figure}

To check this conjecture we ran SIS models on subgraphs containing only the
vertices belonging to either the maximum $k$-core or the star graph centered on the
most connected vertex of the network with degree $\kmax \approx \sqrt{N}$.
Figure~\ref{fig:compare_sig_beta_kcore225and270}(a) shows that the
SIS-$\modstan$ and SIS-$\modall$ essentially  have the same activation threshold for
the maximal $k$-core for both values of $\gamma=2.25$ and $2.7$ while the
activation of the stars centered on the most connected vertex happens in very
different thresholds for these models. The same analysis holds in the not
shown data for SIS-$\modthre$. Therefore, the following framework can be drawn.
For $\gamma=2.25$, the $k$-core is activated first than hubs and the epidemic
activation  is  triggered in the maximal $k$-core for all models. For $\gamma=2.7$,
the hubs are activated firstly for SIS-$\modstan$ while $k$-core is activated
firstly in the other models such that the epidemic activation is due to hubs for
the standard model and still $k$-core for the modified dynamics.

\begin{figure}[hbt!]
	\centering
	\includegraphics[width=0.90\linewidth]{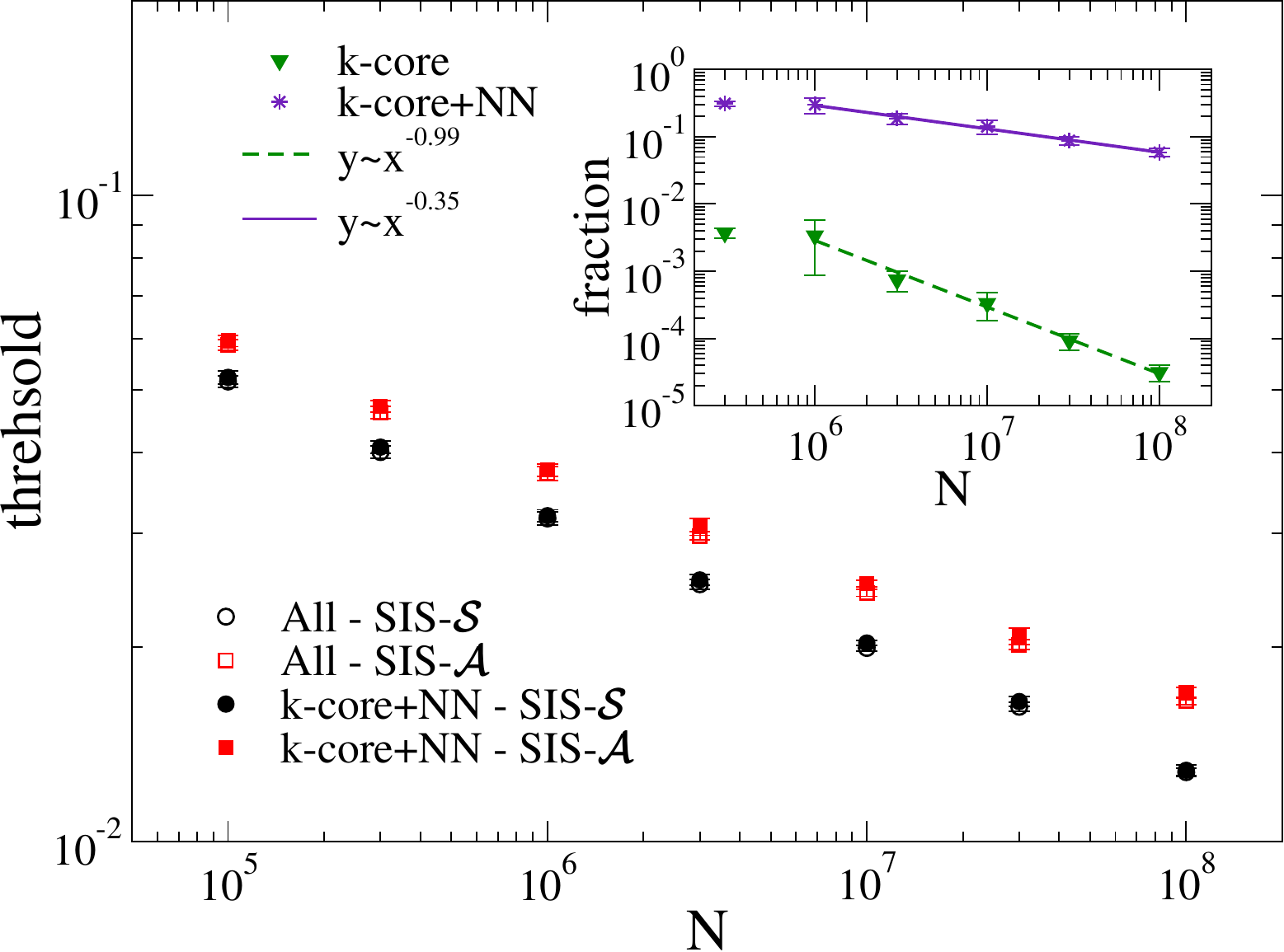}
	\caption{
		Epidemic thresholds for SIS-$\modstan$ and SIS-$\modall$
		running on a subgraph with the maximum $k$-core plus the nearest-neighbor (NN)
		vertices of a UCM network with $\gamma=2.7$.  Inset shows the fraction of the
		network that belongs to the maximal $k$-core including  or not its NNs. Lines
		are power-law regressions.}
	\label{fig:compare_sig_beta_kcoreNN270}
\end{figure}

For $\gamma=2.7$, 
the effective epidemic thresholds for the
entire networks are smaller than those calculated using only the maximal
$k$-core or star centered on the largest hub even with these subgraphs being
associated with the activation of the epidemics. We performed simulations in a
subgraph with the maximal $k$-core plus their nearest-neighbors, which still
represents a sub-extensive fraction of the network as shown in the inset of
Fig.~\ref{fig:compare_sig_beta_kcoreNN270}. The epidemic thresholds in this
subset are essentially the same as those of the whole network for all models, as
shown in Fig.~\ref{fig:compare_sig_beta_kcoreNN270} for SIS-$\modall$ and
SIS-$\modstan$. The trimming of edges reduces the epidemic activity in the
subset containing only the maximal $k$-core while  the $k$-core mediates the
mutual interactions among hubs in the activation driven by them. We see that a
large fraction of the network is redundant for the epidemic threshold
independently if hub (SIS-$\modstan$) or $k$-core activation (SIS-$\modall$ and
$\modthre$) is at work. In both cases, the relevant region to reproduce the
numerical threshold includes the maximal $k$-core plus its nearest-neighbors.

\begin{figure*}[htb]
	\centering
	\includegraphics[width=0.85\linewidth]{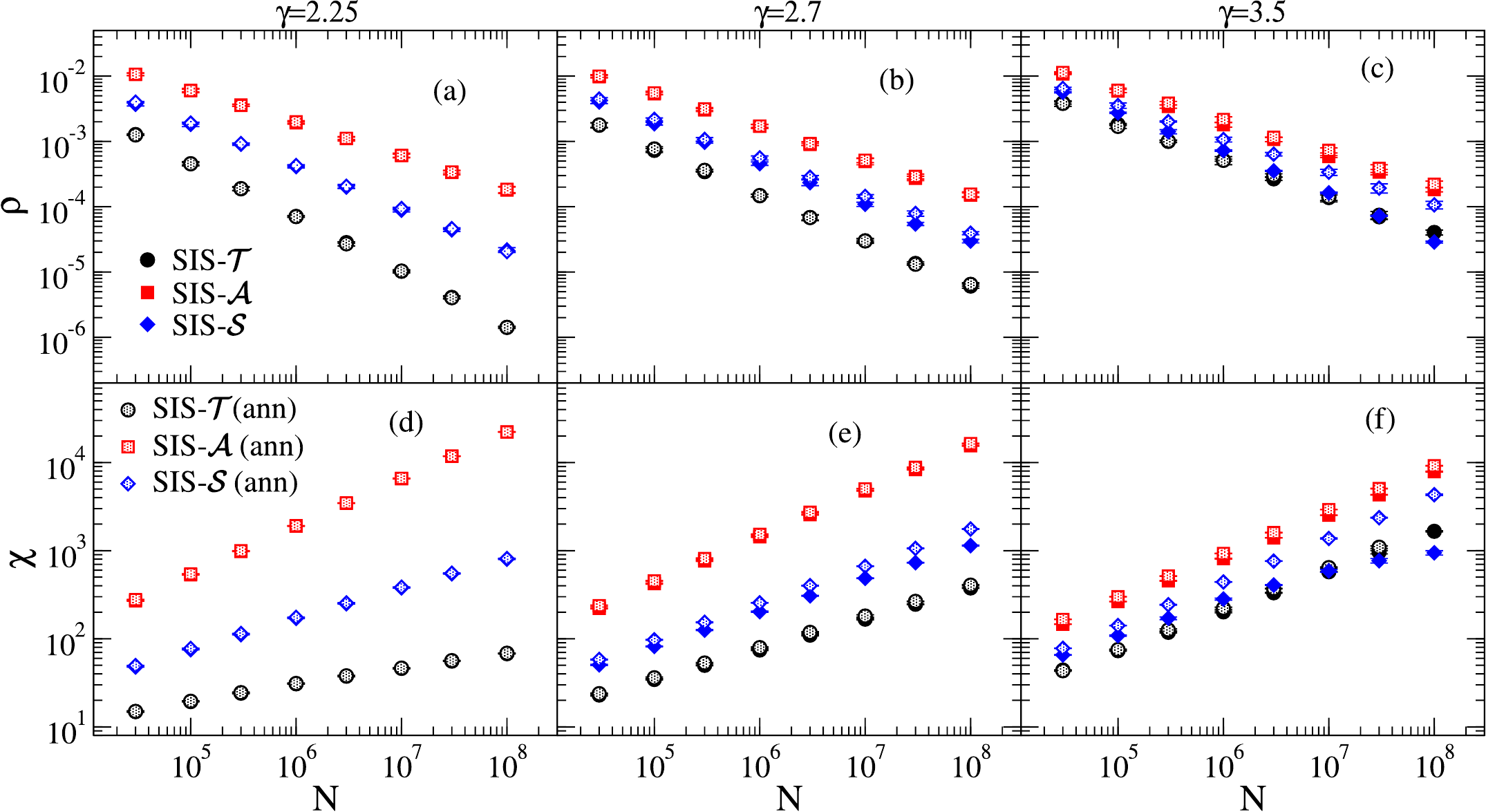}
	\caption[ht]{Finite-size scaling of the critical QS quantities for SIS models
		on UCM networks with different degree exponents. The QS
		densities of infected vertices are shown in
		(a)-(c) while the QS susceptibilities are shown in (d)-(f). The data correspond
		to averages over 10 network realizations and error bars are smaller
		than symbols.}
	\label{fig:fss_sis}
\end{figure*}

Returning to the case $\gamma>3$, UCM networks do not present a  $k$-core structure in
the sense that the decomposition provides a single component containing the
whole network~\cite{Dorogovtsev2006a}. So, since hubs cannot sustain activity
for $\lambda\ll \mu$, the phase transition happens collectively involving a
finite fraction of the network, at a finite threshold~\cite{sander2016}. 

\section{Finite-size scaling of critical quantities}
\label{sec:fss}

The transition between  endemic and disease-free phases can be suited as an
absorbing state phase transition~\cite{Marrobook,henkel2008non}. The finite-size
scaling (FSS) at the critical point (or epidemic threshold) is fundamental for
the characterization of the transition and its critical
exponents~\cite{Marrobook,henkel2008non}. {Several studies concerned with the
universality of the phase transition of the contact process~\cite{Marrobook} 
on complex networks have been performed both numerically and
analytically~\cite{Castellano:2006,Castellano08,Hong2007,Boguna09,Noh2009,
	Ferreira2011,cpannealed,Mata14,RonanEPJB}.} For SIS-$\modstan$, numerical
analyses have been done~\cite{Ferreira12,SanderQS}. A basic approach is to fit
the critical QS density and susceptibility to power-laws in the forms
\begin{equation}
{\rho}\sim N^{-\nu}
\end{equation}
and
\begin{equation}
\chi\sim N^{\phi},
\end{equation}
where $\nu$ and $\phi$ are the critical exponents related to FSS.

\begin{table}[ht]
	\centering
	\def\arraystretch{1.5}
	\caption{Critical exponents  of the FSS for the SIS models on UCM ($\nu$ and $\phi$) and 
		annealed ($\nu_\text{ann}$ and $\phi_\text{ann}$) networks.
		Exponents for SIS-$\modstan$ with $\gamma=3.5$ are missing due to 
		the smearing of the transition.}
	\label{tab:expo}
	\begin{tabular*}{\linewidth}{l @{\extracolsep{\fill}} ccccccc}
		\toprule
		Model &	\multicolumn{2}{c}{$\gamma=2.25$} & \multicolumn{2}{c}{$\gamma=2.7$} & \multicolumn{2}{c}{$\gamma=3.5$} \\\hline
		& $\nu$ & $\nu_\text{ann}$ & $\nu$ & $\nu_\text{ann}$ & $\nu$ & $\nu_\text{ann}$   \\\hline
		$\modthre$&  $0.845(6)$ &$0.84(2)$ & $0.697(4)$ & $0.692(6)$ & $0.55(1)$ & $0.555(3)$  \\ 
		$\modall$ & $0.519(9)$ & $0.517(4)$ & $0.52(1)$ & $0.515(9)$ & $0.499(6)$ & $0.49(3)$ \\
		$\modstan$& $0.63(2)$ & $0.655(2)$ & $0.60(2)$ & $0.57(1)$ & --- &$0.506(7)$  \\\hline
		& $\phi$ & $\phi_\text{ann}$ & $\phi$ & $\phi_\text{ann}$ & $\phi$ & $\phi_\text{ann}$   \\\hline
		$\modthre$ &$0.167(2)$ & $0.169(1)$ & $0.353(1)$ & $0.352(1)$ & $0.458(1)$ & $0.467(3)$ \\
		$\modall$  & $0.530(2)$ & $0.528(2)$ & $0.514(1)$ & $0.513(1)$ & $0.494(1)$ & $0.497(1)$ \\
		$\modstan$& $0.329(5)$ & $0.329(4)$ & $0.372(1)$ & $0.421(1)$ & --- & $0.496(1)$\\	        	
		\toprule
	\end{tabular*}
\end{table}

We considered simulations on annealed networks with same degree distributions as
the quenched ones to represent the mean-field counterpart; see
Appendix~\ref{app:simu} for algorithms. Figure~\ref{fig:fss_sis} presents the
FSS of $\rho$  and $\chi$ at the effective, size-dependent epidemic threshold of
the three SIS models on both UCM and annealed networks. For $\gamma=3.5$, we
used a hard cutoff  $k_c\sim N^{1/\gamma}$ that prevents outliers in the degree
distribution and multiple peaks in the susceptibility curves of quenched
networks~\cite{Ferreira12,Mata2015} making, thus, the determination of the
transition point much more accurate; see Refs.~\cite{Cota,Mata2015} for further
discussion. For $\gamma<3$ the structural cutoff $k_c=\sqrt{N}$ was used.  The
FSS exponents obtained by simple power-law regressions for $N\ge  10^6$ are
shown in Table~\ref{tab:expo}. Uncertainties were calculated using different fit
regions aiming at establishing equivalences or discrepancies between annealed and
quenched simulations rather than accurate estimates of the asymptotic exponents.

The FSS  of the critical quantities provides a scenario in consonance with that
observed for the thresholds. The FSS of both SIS-$\modthre$ and $\modall$ are in
full agreement with the annealed simulations showing their mean-field behaviors
for all values of $\gamma$ investigated. Moreover, the agreement between
quenched and annealed  networks is also found for SIS-$\modstan$ for $\gamma<5/2$.
For $\gamma>5/2$, the dichotomy with respect to SIS-$\modstan$ is again present.
A significant difference in the scaling happens for $\gamma=2.7$ and a sharp
difference is obtained for $\gamma=3.5$. In the latter, we can see a
susceptibility of the SIS-$\modstan$ bending downwardly for the quenched
network, which has been associated to a smearing of the phase
transition~\cite{Cota2016}, while in the annealed network a power-law typical of
an ordinary critical phase transition is seen. No sign of smearing is  observed
for SIS-$\modthre$ and $\modall$.

The FSS provides different exponents for distinct models. So, despite being
described by the same mean-field equations, the role played by stochastic
fluctuations depends on the model. Further analytical studies are required to
clarify the distinction among the exponents.

\section{Discussion}
\label{sec:conclu}

Conception of theoretical frameworks for epidemic processes frequently passes
over the model's fine-tuning due to the belief that universality takes over and
all central features, related to the leading properties and symmetries of a
system, will be obtained irrespective of the specific details. However, this
does not seem to be always the case when the substrate carrying out the process
is a complex network. The standard SIS model, called  SIS-$\modstan$ in this
work, is an example that behaves very differently from most of other related
processes. For example, while many fundamental models on random networks  with a
power-law degree distribution (susceptible infected recovered (SIR)
model~\cite{Moreno2002}, Ising model~\cite{Dorogovtsev2002,Herrero2004},
synchronization~\cite{Rodrigues2016}, etc) have a finite order parameter  for
any value of the control parameter  only for degree exponent
$\gamma<3$~\cite{barratbook}, this happens for any value of $\gamma$ in
SIS-$\modstan$~\cite{Chatterjee09,Castellano2010}. A dichotomy also appears in the two
basic mean-field theories for SIS-$\modstan$, namely, QMF and HMF, which predict
different outcomes for the epidemic threshold for
$\gamma>5/2$~\cite{Castellano2010,Ferreira12}, being only QMF in agreement with
the asymptotically null threshold for $\gamma>3$.

One could naturally wonder if these peculiar characteristics of SIS-$\modstan$
are universal features observed in many other processes.
We investigated two slightly different versions of the
standard SIS,  termed SIS-$\modall$ and SIS-$\modthre$, in which the spontaneous
healing  
and the {unlimited} infection capacity of
a  vertex
are preserved. These
alternative models present exactly the same thresholds of the SIS-$\modstan$ in
both QMF and HMF theories. Stochastic simulations on uncorrelated synthetic
networks, however, show a dual scenario where the three models have essentially
the same vanishing thresholds for $\gamma<5/2$ but disparate results are found
for $\gamma>5/2$. In particular, a finite threshold is observed for $\gamma>3$
in both modified models, in contrast with the asymptotically null threshold of
the standard case. This same framework was observed for SIRS model in
Ref.~\cite{sander2016}, in which an individual acquires temporary immunity when
the agents cannot neither transmit infection nor be infected. The dissonance is
explained in terms of self-sustained, long-lived activation of hubs for any
finite value of $\lambda$~\cite{Boguna2013} that holds for SIS-$\modstan$ but
does not for the other models. The epidemic lifespan of hubs with the modified
dynamics increases slowly (algebraically or logarithmically) with the hub degree
in contrast with the exponential increase of the standard case. The last one
permits the
long-range mutual activation or reactivation of hubs~\cite{Boguna2013,sander2016}.

We also analyzed the activation mechanisms of the epidemic phase on uncorrelated
networks. While the activation for SIS-$\modstan$ occurs in the innermost,
densely connected core of the network, determined by the largest index of a
$k$-core decomposition, for $\gamma<5/2$ and in hubs for
$\gamma>5/2$~\cite{Castellano12}, this happens for the whole range of scale-free
networks with $2<\gamma<3$ for SIS-$\modall$ and SIS-$\modthre$. Absence of a
$k$-core organization~\cite{Dorogovtsev2006a} and a short-lived  activity in
star subgraphs as $\lambda\rightarrow 0$ for $\gamma>3$ suggests that the
activation of the epidemic phase in the modified SIS models is collective,
involving essentially the whole network~\cite{sander2016}, and occurs at a
finite threshold.

The aforementioned dichotomy is also observed in the finite-size scaling of the
quasi-stationary density and susceptibility computed at the epidemic threshold.
Agreements between simulations on quenched and annealed versions of
the investigated networks are observed for SIS-$\modall$ and SIS-$\modthre$
irrespective of the degree exponent. In turn, they deviate in
the hub activated regime with $\gamma>5/2$ in SIS-$\modstan$, being more marked
for $\gamma>3$ where the transition observed for quenched networks seems to be
smeared~\cite{Cota2016}, in contrast with a regular critical transition in the
annealed case.

Here, we also comment the nature of the epidemic activation in processes with
spontaneous  healing with uniform  rates and a bounded infection produced by a
vertex, differing from the three SIS models investigated here and from
SIRS~\cite{sander2016}. In these bounded infection models, the epidemic lifespan
on stars is finite for any value of the infection rate~\cite{sander2016} and the
epidemics can be activated only collectively in a finite threshold for any value
of $\gamma$, as observed in simulations of the  contact process on quenched
networks ~\cite{Castellano:2006,Hong2007,Ferreira2011}, for example.
Table~\ref{tab:activ_mech} summarizes the activation mechanism of the  different
epidemic models investigated or discussed in the present work.

\begin{table}[ht]
	\centering
	\def\arraystretch{1.5}
	\caption{Activation mechanisms for different epidemic models presenting active
		steady states on uncorrelated networks with degree distribution
		$P(k)\sim k^{-\gamma}$.}
	\label{tab:activ_mech}
	\begin{tabular*}{\linewidth}{l @{\extracolsep{\fill}} ccc}
		\toprule
		Model          & $2<\gamma<5/2$ & $5/2<\gamma<3$ & $\gamma>3$ \\\hline
		SIS-$\modstan$ &  Max $k$-core  &  Hub           &  Hub \\
		SIS-$\modthre$ &  Max $k$-core  &  Max $k$-core  &  Collective\\
		SIS-$\modall$  &  Max $k$-core  &  Max $k$-core  &  Collective\\
		SIRS           &  Max $k$-core  &  Max $k$-core  &  Collective\\
		CP&  Collective    &  Collective    &  Collective\\
		\toprule
	\end{tabular*}
\end{table}

{An interesting} point observed in our analysis  is
that the 
HMF theory was  more accurate than 
QMF theory in all investigated cases, except for SIS-$\modstan$. 
Dynamical correlations 
are neglected in both approaches assuming that the states of interacting
vertices, in case of QMF, or interacting compartments, in the case of HMF, are
independent. This approximation becomes more problematic for QMF since we
explicitly reckon the interactions with the actual nearest-neighbors of each
vertex and
assume that their states are independent. The
leading approximation in HMF is to assume that the probability to be infected
depends only on the vertex degree, neglecting the local structure of the
network. As an effect, HMF theory may  not be able to capture localized activity
due to specific motifs as those observed for star subgraphs in the
SIS-$\modstan$ model. 
Finally, QMF theory is not a genuine mean-field approach since it does not
present mixing of vertices while HMF does through the degree
compartmentalization. Our results thus reinforces the belief that mean-field
approaches with heterogeneous mixing are suitable approximations for most
dynamical processes on networks with a small-world property, in which the
average distance between vertices increases logarithmically with the system
size~\cite{barabasi2016network}.

Our results gathered with previous reports raise an important question on the
modeling of epidemic processes on networks. Once details may matter, which would
be the actual mechanisms used in models that correspond to real epidemics and
which would be the best approaches to analytically investigate real epidemic
processes? The summary presented in Table~\ref{tab:activ_mech} suggests that the
hub activation mechanism, intensively investigated
recently~\cite{Chatterjee09,Castellano2010,Goltsev2012,St-Onge2017,Wei2017,Boguna2013,Lee2013,Mata2015,sander2016}, seems to be more a peculiarity than a rule in epidemic spreading. {We expect that our results will guide the analysis of other classes of the dynamical process on networks such as the complex contagion models~\cite{Granovetter1978,Centola2007,Campbell2013,Karsai2014} where activation requires more than one interaction to be effected. }

\begin{acknowledgments}
	This work was partially supported by the Brazilian agencies CAPES, CNPq, and
	FAPEMIG. We thank the support from the program  \textit{Ci\^encia sem
		Fronteiras} - CAPES under project No. 88881.030375/2013-01.
	
	This is a preprint version of the published article at \href{https://journals.aps.org/pre/abstract/10.1103/PhysRevE.98.012310}{Physical Review E \textbf{98}, 012310 (2018)}.
\end{acknowledgments}

\appendix

\section{Computer implementations of the epidemic models}
\label{app:simu}

To build the computer implementations, all involved rates are reckoned using
statistically exact prescriptions based on the Gillespie
algorithms~\cite{Gillespie1976403}. We consider phantom processes that do
nothing but counting for time increments. These ideas are detailed in
Ref.~\cite{Cota}. Below we present recipes for the  models investigated in the
present work.

\subsection{SIS-$\modstan$}
\label{app:sigma}
The SIS-$\modstan$ dynamics in a network of size $N$ with infection and healing
rates $\lambda$  and $\mu$ can  be efficiently simulated as follows. A list with
all infected vertices, their number $\Ninf$, and the number of edges $\Ne$
emanating from them are recorded and constantly updated. Each time step involves
the following procedures. (i) With probability
\begin{equation}
p = \frac{\mu \Ninf}{\mu \Ninf+\lambda \Ne},
\end{equation}
an infected vertex is selected with equal chance and healed.
(ii) With complementary probability $1 - p$, an infected vertex is selected with
probability proportional to its degree. A neighbor of the selected vertex is
chosen with equal chance and, if susceptible, is infected. Otherwise, no change
of state is implemented (it is a phantom process). (iii) The time is incremented by
\begin{equation}
\tau = \frac{-\ln(u)}{\mu \Ninf + \lambda \Ne},
\end{equation}
where $u$ is a pseudo random number uniformly distributed in the interval $(0,1)$
and the simulation runs to the next step.

\subsection{SIS-$\modall$}
\label{app:beta}

This model implementation is very similar to the contact
process~\cite{Marrobook}. A list with the infected vertices and their
number $\Ninf$ is built and constantly updated.
At each time step, the rules  are the following. (i)  With probability
\begin{equation}
p =\frac{\mu}{\mu+\lambda},
\end{equation}
an infected vertex is randomly chosen  and healed. (ii) With complementary
probability $1-p$, all susceptible neighbors of a randomly chosen infected
vertex are infected at once. (iii) The time is incremented by
\begin{equation}
\tau = \frac{-\ln(u)}{(\mu +\lambda)\Ninf}.
\end{equation}

\subsection{SIS-$\modthre$}
\label{app:alpha}
As in SIS-$\modstan$ and $\modall$, a list containing the infected vertices and their number
$\Ninf$ is built and constantly updated. We have also to maintain  an auxiliary
list including the number of infected neighbors $n_i$ of each vertex $i$ and the
total number of susceptible vertices $\NSI$ that have at least one infected
neighbor. At each time step, the rules  are the following.  (i) With probability
 \begin{equation}
 p=\frac{\mu \Ninf}{\mu\Ninf+\lambda \NSI},
 \end{equation}
an infected vertex is selected with equal chance and healed.
(ii) With complementary probability $1 - p$, an infected vertex is selected with
probability proportional to its degree and one of its neighbors is randomly chosen.
If the selected  neighbor $i$ is susceptible it is accepted and infected with
probability $1/n_i$. The procedure of choosing a susceptible vertex
 is repeated until one of them is
found. (iii) The time is incremented by
\begin{equation}
\tau = \frac{-\ln(u)}{\mu \Ninf +\lambda \NSI}.
\end{equation}

\subsection{Simulation on uncorrelated annealed networks}

On uncorrelated annealed networks, the unique difference in SIS-$\modstan$ and
SIS-$\modall$ with respect to the quenched case is that the choice of the
neighbors to be infected is done by selecting any vertex of the network with
probability proportional to its degree.

For SIS-$\modthre$, however, the algorithm becomes trickier and, consequently, very slow. 
The probability that a susceptible 
vertex $j$ becomes infected is given by 
\begin{equation}
P_j=1-(1-\Theta)^{k_j},
\end{equation}
where $k_j$ is the degree of vertex $j$ and $\Theta = \Ne/(N\av{k})$ is the
probability that a randomly selected neighbor (at the other side of the edge) is
infected in the annealed  network. Let us define a total rate that one tries to
infect a susceptible vertex as $L=\lambda (N-\Ninf)$, which is larger than the
real one since only the susceptible vertices that have at least one infected
neighbor can actually be infected and this happens with probability $P_j<1$. The
total rate that a vertex is healed is $M=\mu \Ninf$. The algorithm becomes the
following. (i) An infected vertex is randomly chosen and healed with probability
$p=M/(L+M)$. (ii) With probability $1-p$, a susceptible vertex is randomly chosen
and infected with probability $P_j$. (iii) The time is incremented by $\tau =
-\ln(u)/(L+M)$.

The exactness of these algorithms is confirmed in Fig.~\ref{fig:hmfanng3p5}
where simulations on annealed networks are compared with the integration of 
the HMF equations.

\section{Pairwise approximations for SIS-$\modstan$}
\label{app:pair}

\begin{figure}[H]
\centering
\includegraphics[width=0.90\linewidth]{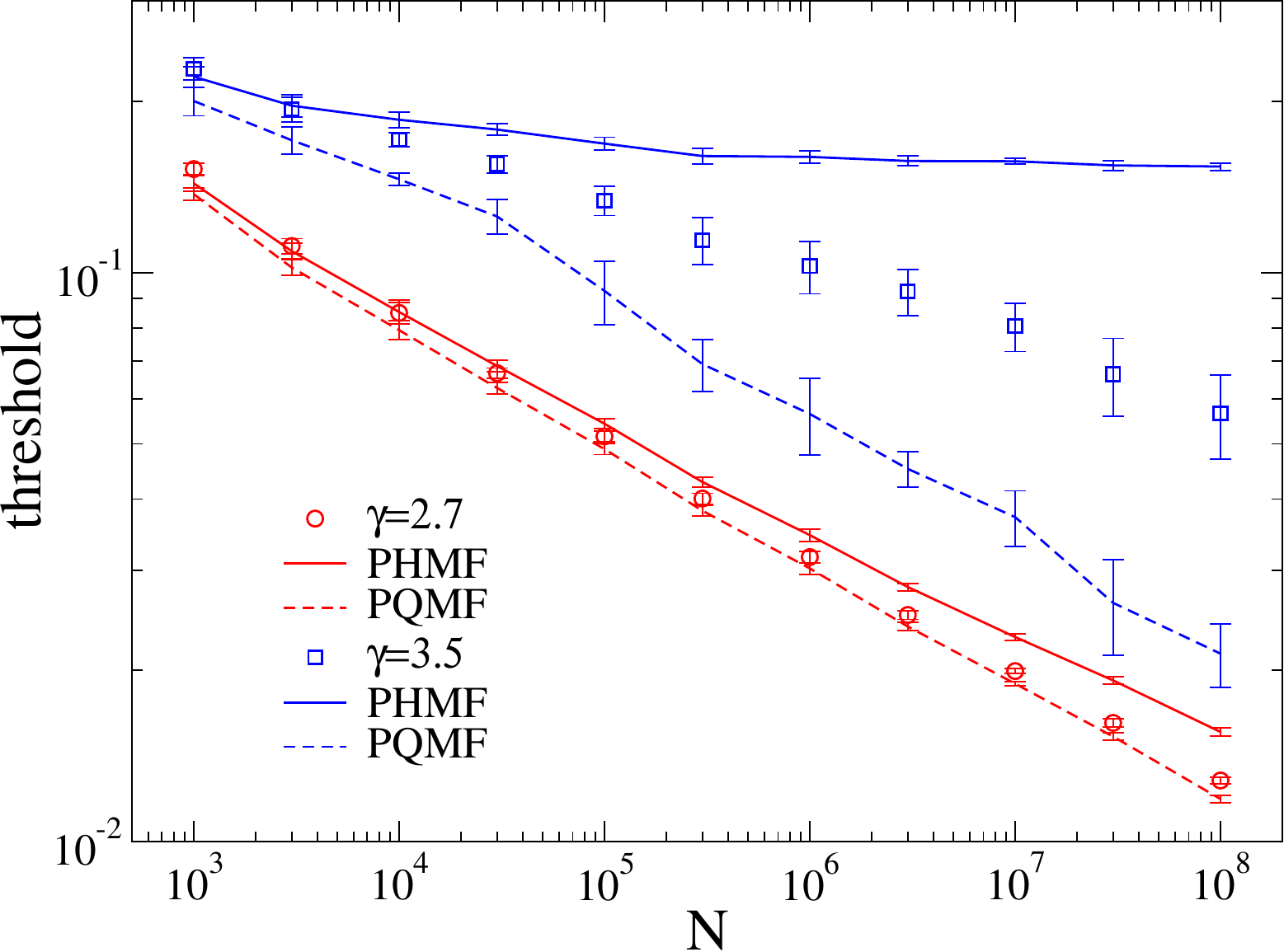}
\caption{Comparison of pairwise approximations  with simulations (symbols) for
	epidemic thresholds of the SIS-$\modstan$ on UCM networks, given by
	Eqs.~\eqref{eq:lbcPHMF} 	 and \eqref{eq:lbcQHMF}. Data correspond to averages
	over 10 network samples.}
\label{fig:phmf_pqmf_ucm}
\end{figure}

The pairwise heterogeneous mean-field (PHMF) approximation for 
SIS-$\modstan$ with $\mu=1$ on uncorrelated networks provides a threshold~\cite{Mata14} 
\begin{equation}
\lambda_c^\text{PHMF} = \frac{\av{k}}{\av{k^2}-\av{k}}.
\label{eq:lbcPHMF}
\end{equation}
The threshold of the pairwise quenched mean-field approximation (PQMF)
is obtained when the largest eigenvalue of the matrix~\cite{mata2013pair},
\begin{equation}
L_{ij}=-\left(1+\frac{\lambda^2 k_i}{2\lambda+2}\right)\delta_{ij} + 
\frac{\lambda(2+\lambda)}{2\lambda+2}A_{ij},
\label{eq:lbcQHMF}
\end{equation}
is null. Figure~\ref{fig:phmf_pqmf_ucm} shows the thresholds of the pairwise
theories computed for UCM networks.

\section{Approximated expressions for epidemic lifespan on star graphs}
\label{app:lsanalytic}

To obtain approximated expressions for the  lifespan of the SIS epidemic
processes ($\modstan$, $\modall$, and $\modthre$) on a star graph with $k$ leaves,  
we consider the following discrete time dynamics based in Ref.~\cite{sander2016}:

(i) At time $t=0$, the center is infected and all leaves are susceptible.

(ii) At a time $t=t_1$, the center is healed and $n$ leaves are simultaneously 
 infected with probability $P_1(n|k)$.

(iii) At time $t=t_1+t_2$, the center is reinfected and all leaves become simultaneously 
susceptible. This occurs with probability $P_2(n)$.

The probability that the dynamics survives after this sequence is
\begin{equation}
Q=\sum_{n=1}^{k} P_2(n) P_1(n|k),
\label{eq:Qger}
\end{equation}
and the probability that the dynamics ends up at the $s$th step
 is $Q^{s-1} (1-Q)$. So, the average number of steps is 
\begin{equation}
	\av{s}=\sum_{s=0}^{\infty}sQ^{s-1} (1-Q)=\frac{1}{1-Q}.
	\label{eq:avsger}
\end{equation}
Next, we define the times $t_i$ and probabilities $P_i$ ($i=1,2$) for each model.

The steps for standard SIS~\cite{Boguna2013,sander2016} are reproduced here as a
guide to the other models. We chose $t_1=t_2={1}/{\mu}$, which is the average
time that a vertex takes to be healed. The probability that the center
infects a leaf before healing is $p=\lambda/(\mu+\lambda)$~\cite{Boguna2013},
which is the same for all leaves. So, the probability that $n$ leaves were
infected at time $t_1$ becomes
\begin{equation}
P_1(n|k)=\binom{k}{n}p^n(1-p)^{k-n}.
\label{eq:P1sigma}
\end{equation}
The probability that at least one leaf reinfects the center before  healing at time $t_2$ 
is
\begin{equation}
P_2(n)=1-(1-p)^n.
\label{eq:P2sigma}
\end{equation}
Plugging Eqs.~\eqref{eq:P1sigma} and \eqref{eq:P2sigma} in \eqref{eq:Qger}
we obtain
\begin{equation}
Q=1-(1-p^2)^k\approx 1-\exp(-k\lambda^2/\mu^2)
\label{eq:Qsigma}
\end{equation}
where the approximation holds for the regime $\lambda\ll \mu$, in which were are
interested in. Now, substituting Eq.~\eqref{eq:Qsigma} into Eq.~\eqref{eq:avsger}, we obtain
\begin{equation}
\tau_k^{\modstan}=(t_1+t_2)\av{s}\approx \frac{2}{\mu}\exp\left(\frac{\lambda^2}{\mu^2}k\right).
\end{equation}
The prediction is an exponential increase with the star size.

For SIS-$\modall$, since all leaves are simultaneously infected before healing
with probability $p=\lambda/(\mu+\lambda)$ we have that $P_1(n|k)=p\delta_{n,k}$
and the other variables are assumed to be the same. So, we have $Q=p[1-(1-p)^k]$
which leads to $\av{s}\approx 1$ and the epidemic lifespan
\begin{equation}
\tau_k^\modall =  (t_1+t_2)\av{s} \approx \frac{2}{\mu}
\end{equation}
for $\lambda\ll\mu$. The prediction is a finite lifespan. 

For SIS-$\modthre$ we have the same expression of SIS-$\modstan$ for $P_1(n|k)$
while the probability that center is reinfected is simply $P_2(n)=p$,
irrespective of $n$. So, $\av{s}\approx 1$ as in SIS-$\modall$.
However, since  infection rate of the center is independent of how many infected
leaves are present,  we must use the average time for all leaves to be healed
instead of the average time for a single leaf to be healed. Considering the
healing processes of each leaf as being an  independent Poisson process and neglecting the
possibility of reinfections of leaves during this process, the average time for
$n$ leaves to be healed is
\begin{equation}
t_2^{(n)}=\int_{0}^{\infty} t \left[n \left(1-e^{-\mu t}\right)^{n-1} e^{-\mu t}\right] \mu  dt \approx \frac{0.92}{\mu} \ln n.
\end{equation}
The term between brackets is the probability that one single leaf is infected
at time $t$, $\mu dt$ is the probability that it heals at time $t$, and the
saddle point approximation was used to compute the integral assuming $n\gg 1$.
So, replacing $n$ by the average number of infected leaves in part (ii), $\av{n}=pk$,
to estimate $t_2=\av{t_2^{(n)}}\approx t_2^{(\av{n})}$, we obtain
\begin{equation}
\tau_k^\modthre\approx \frac{1+0.92\ln (pk)}{\mu}\simeq\frac{0.92}{\mu}\ln k
\end{equation}
for $\lambda\ll\mu$. The prediction is a logarithmic increase with the star size.


%

\end{document}